\documentclass[12pt,a4paper,final]{iopart}

\usepackage{iopams}
\usepackage{graphicx}
\usepackage{amssymb,amsthm}
\usepackage{amscd}	%produces rectangular diagrams, arrows and equal signs
\usepackage{enumerate}
\usepackage{framed}
\usepackage{subfigure}
\usepackage{xcolor}
\usepackage{color}
\usepackage{cases}
\usepackage[margin=1cm]{caption}

\newcommand{\bra}[1]{\mbox{$\left\langle #1 \right|$}}
\newcommand{\ket}[1]{\mbox{$\left| #1 \right\rangle$}}
\newcommand{\braket}[2]{\mbox{$\left\langle #1 | #2 \right\rangle$}}

\newcommand{\eqref}[1]{(\ref{#1})}
\newtheorem{theorem}{Theorem}
\eqnobysec

\begin{document}

%\title[Preparing an article for IOP journals in  \LaTeXe]{Preparing an article for publication in an Institute of Physics Publishing journal using \LaTeX}

%\author{Author One$^{1,2}$}
%\address{$^1$Address One, Neverland}
%\address{$^2$Address Two, Neverland}
%\ead{author.one@mail.com}

%\author{Author Two}
%\address{Address Three, Neverland}
%\ead{author.two@mail.com}

%\author[cor1]{Author Three}
%\address{Address Four, Neverland}
%\eads{\mailto{author.three@mail.com}, \mailto{author.three@gmail.com}}

% title

\title{Loss-tolerant measurement-device-independent quantum random number generation}

\author{Zhu Cao, Hongyi Zhou, Xiongfeng Ma}
\address{Center for Quantum Information, Institute for Interdisciplinary Information Sciences, Tsinghua University, Beijing, China}
%\email{caozhu55@gmail.com}
%\email{xma@tsinghua.edu.cn}

\begin{abstract}
Quantum random number generators (QRNGs) output genuine random numbers based upon the uncertainty principle. A QRNG contains two parts in general --- a randomness source and a readout detector. How to remove detector imperfections has been one of the most important questions in practical randomness generation. We propose a simple solution, measurement-device-independent QRNG, which not only removes all detector side channels but is robust against losses. In contrast to previous fully device-independent QRNGs, our scheme does not require high detector efficiency or nonlocality tests. Simulations show that our protocol can be implemented efficiently with a practical coherent state laser and other standard optical components. The security analysis of our QRNG consists mainly of two parts: measurement tomography and randomness quantification, where several new techniques are developed to characterize the randomness associated with a positive-operator valued measure.
\end{abstract}

\pacs{}
\vspace{2pc}
%\submitto{\NJP}
\maketitle

\section{Introduction}
Random numbers have applications in many fields including industry, scientific computing, and cryptography \cite{eastlake2005s,brown1994security}. In particular, the randomness of the key is the security foundation for all the cryptographic tasks. Any bias on random numbers may result in security loopholes \cite{Shannon:1949:OTP}.

%Especially, in secure communications, such quantum key distribution \cite{Bennett:BB84:1984}, a significant branch of quantum cryptography, the randomness of the key is the security foundation of communication protocols.

Traditionally, there are two types of random number generators (RNGs), pseudo-RNGs and physical RNGs. A pseudo-RNG is a deterministic expansion of random seeds and hence not random \cite{knuth2014art}. A physical RNG is based on chaotic physical process such as noise in electric devices \cite{jun1999intel}, oscillator jitter \cite{bucci2003high}, and circuit decay \cite{tokunaga2008true}. Since a full characterization of a physical RNG process may enable an adversary to predict the outcomes, the randomness is not information-theoretically provable. In practice, it is very challenging to rule out the bias in output random numbers, and hence these physical RNGs may lead to security loopholes when employed in cryptographic tasks.

On the other hand, quantum random number generators (QRNGs), stemming from the intrinsic uncertainty of quantum measurement outcomes, are able to output randomness that is guaranteed by quantum mechanics. Some popular QRNG schemes include single photon detection \cite{Dynes:QRNG:2008,wayne2010low,furst2010high}, vacuum state fluctuation \cite{Gabriel:QRNG:2010} and quantum phase fluctuation \cite{Qi2010QRNG,Xu:QRNG:2012}. The output randomness of these QRNG relies on assumptions on the realization devices. In practice, however, device imperfections may lead to potential loopholes, which can be exploited by an adversary.

To solve this problem, device-independent QRNG (DIQRNG) schemes, whose output randomness does not rely on specific physical implementations, have been proposed \cite{Colbeck09}. Based on quantum non-locality, such a DIQRNG is mainly designed with entangled particles and can certify genuine randomness. By performing measurements on two entangled systems and checking whether the correlation violates a certain Bell inequality, true random numbers are generated. It has been proved that high detection efficiency (over $2/3$) and space separation are necessary in such a device-independent scheme \cite{eberhard1993background,massar2003violation}. However, normal optical detectors, with which all practical fast QRNGs are built, only have an overall efficiency around $10\%$ and do not satisfy this condition.  In fact, loophole-free DIQRNGs have not yet been demonstrated in labs up till now \cite{PhysRevLett.111.130406}.
%, which has only been achieved in ion trap systems and photon detection with transition-edge super-conducting sensors so far.

%attack are targeted at measurement device, which is the motivation of MDI QRNG
Similar issues also exist in another quantum cryptographic task --- quantum key distribution (QKD). In order to solve the practical issues in the device-independent schemes, additional assumptions are added to make the schemes more practical \cite{Ma2012DDIQKD,Pawlowski:Semi:2011}. In particular, a measurement-device-independent (MDI) QKD scheme is proposed \cite{Lo:MDIQKD:2012} such that all the detection loopholes can be removed using trusted source devices. The MDIQKD scheme turns out to be loss-tolerant and very effective to defend against practical attacks \cite{PhysRevA.74.022313,Qi2007Time}, without using complicated characterization on devices \cite{fung2009security}. The security of MDIQKD stems from the time-reversed EPR-based QKD protocols \cite{Biham:1996:Quantum,Inamori:TimeReverseEPR:2002,Stefano:MDIQKD:2012}.

Unfortunately, the idea of MDIQKD cannot directly apply to the task of QRNG due to the subtle difference between QKD and QRNG in practice. In QKD, local randomness is assumed to be a free resource, while in QRNG, (local) randomness is the goal to pursue. In fact, the randomness generated by the measurement (at most 2 bit per run) is less than the randomness required for the state preparations (4 bit per run) in MDIQKD \cite{Lo:MDIQKD:2012}. Intuitively, the measurement in MDIQKD only establishes correlation between the two communication parties and helps to generate a shared randomness, but it does not generate additional randomness.

Recently, there are a few attempts that tackle the challenge of MDI QRNG, including a qubit-modeled QRNG \cite{canas2014experimental} and an MDI entanglement witness (MDIEW) based QRNG \cite{banik2014measurement}. These schemes are more secure than conventional QRNGs, in the sense that some of the assumptions on the devices are removed. Comparing to DIQRNGs, they are more practical on loss-tolerance.
%the still some disadvantages that need to be improved. In the qubit-modeled QRNG, first, both the source and the measurement device are assumed to be qubit systems, which may be unrealistic in practical system; second, preparing a single photon state is difficult to realize in practice; third, it needs input randomness larger than the output randomness.
However, a key assumption in the first scheme \cite{canas2014experimental}, that both the source and the measurement device are assumed to be qubit systems, is difficult to be fulfilled in practice.
 %In the MDIEW based QRNG, quantum communication should be forbidden by a space-like separation, which requires real-time manipulations and synchronized operations and is thus hard for a practical implementation. Further, it cannot tolerate basis-dependent losses, which for example can occur in a quantum dephasing channel.
For the second scheme \cite{banik2014measurement},
%it requires space-like separation of two measurement devices which adds complication to the practical implementation. And moreover,
it cannot tolerate basis-dependent losses, which puts strict constraints on measurement devices.

Here, we present a loss-tolerant MDI QRNG scheme, stemming from a simple qubit scheme that measures a state  $\ket{+}=(\ket{0}+\ket{1})/\sqrt{2}$ in the basis of \{$\ket{0}$, $\ket{1}$\}. The randomness is originated in breaking the coherence of the input state \cite{Yuan2015Coherence}. In order to validate the measurement devices, several additional quantum input states need to be sent. Such validation procedure is related the concept of self-testing \cite{MayersYao_98}. For example, the source could check if the measurement device always outputs the correct eigenstate when inputting the state \ket{0}. Note that if the measurement device faithfully measures in the \{$\ket{0}$, $\ket{1}$\} basis, it should always output $\ket{0}$ deterministically. To reduce the input randomness, testing input states should be rarely sent. In our analysis, we do not require the source to be a single photon source. Instead, practical photon sources, such as a weak coherent state source, can be used in our scheme.

The organization of the paper is as follows. In Section \ref{sec:scheme}, we give a formal description of our protocol.
 In Sections \ref{sec:tomog} to \ref{sec:coh}, we analyze our protocol.  Our protocol can be divided into two parts, measurement tomography and randomness quantification of a POVM, thus Section \ref{sec:tomog} and Section \ref{sec:POVM} are devoted to these two parts respectively. In Section \ref{sec:stat}, we analyze the finite size effect. Section \ref{sec:coh} extends the analysis from a single-photon source to a coherent-state source.
 Finally we conclude in Section \ref{sec:concl}.

\section{Brief description of MDI QRNG} \label{sec:scheme}
In our MDI QRNG scheme, a quantum source emits signals, which is measured by an untrusted and uncharacterized device. The process is repeated for $n$ times, among which some of the runs are chosen as test runs and the rest for randomness generation. In test runs, a measurement tomography is performed, while in a generation run, random numbers are generated. The protocol is presented in Fig.~\ref{fig:pro}.

\begin{table}[htbp]
\caption{The measurement device is designed to measure in the $\sigma_z$ basis for $n$ runs.
 At the end of the protocol, the measurement device outputs a uniformly random string of length $rn$, where $r$ is the product of the ratio for generation runs and the min-entropy of the raw measurement outcomes.}
\begin{framed}
\centering
\begin{enumerate}
\item
\textbf{Random seed:} The user, Alice, randomly chooses a subset $B\subset\{1,\cdots,n\}$ from the $n$ runs. %Here the size of the subset is determined by the security parameters discussed later.

\item
\textbf{Test mode:}
For rounds in the subset $B$, a trusted source randomly emits qubit states $\ket{0},\ket{1}, \ket{+}, \ket{+i}$ to an untrusted measurement device, where $\ket{0},\ket{1}, \ket{+}, \ket{+i}$ are eigenstates of Pauli matrices $\sigma_z, \sigma_z, \sigma_x, \sigma_y$, respectively. Then the measurement device outputs bits $b\in \{0,1\}$. Alice uses these outputs to perform a measurement tomography.

\item
\textbf{Generation mode:} For the runs not in $B$, Alice sends the measurement device a fixed state of $\ket{+}$. Again, the measurement device outputs bits $b\in \{0,1\}$.

\item
\textbf{Extraction:} Randomness extraction is performed on the raw outputs to obtain a uniformly random string of length $rn$. The min-entropy of the raw data is determined by the tomography results.
\end{enumerate}
\end{framed}
\label{fig:pro}
\end{table}

%Intuition why the protocol works.
Here is the intuition why the protocol works. From the test runs, the measurement tomography is used to monitor the devices in real time. If the tomography result passes certain threshold, the user is sure that the measurement devices function properly. Of particular interest is that how the protocol deals with losses in order to make it loss-tolerant. We emphasize that in the protocol we do not discard the loss events. Instead, the measurement device should always output 0 or 1. In practice, if there is no detection click, the measurement device outputs 0. Intuitively, the positions of the loss are mixed with real detected bits 0, restricting the adversary's ability to output a fixed string.

Let us consider a simple attack that works for conventional QRNGs when the measurement devices are untrusted and the loss is over 50\%. A successful attack can be defined as follows: an adversary, Eve, can manipulate the QRNG so that it outputs a predetermined string (which could appear random to Alice)\footnote{This is a classical adversary scenario, which can be extended to quantum adversary scenario \cite{Miller15}.}. When Eve can fully control the measurement devices, she first performs the faithful measurement (without losses) designated by the protocol. Then within the measurement outcomes, Eve post-selects a string according to her predetermined string (which could appear random to Alice). The post-selection works as follows: if a measurement outcome matches the corresponding bit in Eve's predetermined string, Eve announces the outcome, otherwise she announces a loss. Then if the measurement outcomes contain an equal number of 0s and 1s, approximately 50\% of outcomes will be announced as losses. Thus the output string could be predetermined without being noticed by the user.

Such attack will not work for our MDI QRNG. If Eve performs this attack and outputs 0 when she wishes to announce a loss, each bit of the outcomes will now independently have probability $3/4$ to be 0 and $1/4$ to be 1. Thus the randomness of the output is $\log_2 (4/3)$ per bit, which is nonzero.

By the protocol description, the randomness analysis can be naturally decomposed into two parts, measurement tomography and randomness quantification given a known positive-operator-valued measure (POVM). We thus divide the analysis into the following two sections accordingly.

\section{Measurement tomography} \label{sec:tomog}
In this section, we investigate the following question. Given a trusted single photon source, which is treated as a qubit, how to make a measurement tomography on a detection device, whose dimension is unknown? Later, we will discuss how to replace the single photon source with a more practical coherent state source.

Generally, there are three types of attacks for security protocols, individual attack where Eve performs an identical and independent attack on each run, collective attack where Eve probes the input state in each run separately and performs a joint post-processing, and coherent attack where Eve might exploit the correlation between the runs by probing all the inputs jointly \cite{Biham2002Security}. In our protocol, to be more specific, an individual attack means that the POVM of Eve in different runs will be the same; a collective attack means Eve performs different POVMs in different runs but uncorrelated; a coherent attack means the POVMs in different runs are correlated. We will extend our security proof framework from individual attack to collective attack, and leave coherent attack for future research.

Recall that we have restricted the measurement device to always output 1 and 0 in each run. Though the adversary could add an arbitrary number of ancillaries to perform a high-dimensional PVM, its measurement operator can always be described by a two-dimensional POVM with two outcomes $\{F_0, F_1\}$ where $F_0+F_1=I$, because of the qubit input. Here, we start with the analysis under individual attacks and hence we can assume the POVM elements are the same for every run. The extension to collective attacks will be presented in Sec.~\ref{sec:coll}.

For a qubit input state $\rho$, the probabilities of outputting 0 and 1 are given by
\begin{equation} \label{eq:POVMoutcome}
\eqalign{
\textsf{Prob}(0|\rho)=tr(\rho F_0), \\
\textsf{Prob}(1|\rho)=tr(\rho F_1).
}
\end{equation}
Any two-dimensional POVM has the form \cite{povm}
\begin{equation}
\eqalign{
\label{Eq:povm}
F_0=a_1(I+\vec{n}_1\cdot \sigma),\\
F_1=a_2(I+\vec{n}_2\cdot \sigma),}
\end{equation}
where $\sigma$ is the vector composed of three Pauli matrices, $\vec{n}_1=(n_x,n_y,n_z)$ and $\vec{n}_2$ are three-dimensional real number vectors. The coefficients are real numbers and satisfy
\begin{equation}
\eqalign{
\label{eq:coefficientsrelation}
a_1, a_2\ge 0, \quad a_1+a_2=1, \\
|\vec{n}_1|, |\vec{n}_2| \le 1, \quad a_1\vec{n}_1+a_2\vec{n}_2=0.
}
\end{equation}

In measurement tomography, one can input the four basis of two-dimensional density matrices, $(I+\sigma_z)/2$, $(I-\sigma_z)/2$, $(I+\sigma_x)/2$, and $(I+\sigma_y)/2$, which correspond to pure states $\ket{0}$, $\ket{1}$, $\ket{+}$, and $\ket{+i}$, respectively. The probabilities of outputting $0$ for the four states can be estimated through counting the ratio of $0$s in the test runs. When there are an infinite number of runs, the estimation can be done accurately. From Eq.~\eqref{eq:POVMoutcome},
these probabilities are given by
\begin{equation} \label{eq:tom}
\eqalign{
\textsf{Prob}(0|(I+\sigma_z)/2)&= a_1+a_1n_z,  \\
\textsf{Prob}(0|(I-\sigma_z)/2)&= a_1-a_1n_z,  \\
\textsf{Prob}(0|(I+\sigma_x)/2)&= a_1+a_1n_x, \\
\textsf{Prob}(0|(I+\sigma_y)/2)&= a_1+a_1n_y.
}
\end{equation}
Then the coefficients $a_1, n_x, n_y, n_z$ can be solved given the measurement results, the left side quantities of Eq.~\eqref{eq:tom}. Note that if the input is a linear combination of these four inputs, the probability of outputting $0$ will also be a corresponding linear combination of the above four probabilities. Without loss of generality and for ease of discussion, we will assume $a_1\le a_2$ hereafter.

There also exist tomography methods for coherent state source \cite{luis1999complete,d2004quantum,lundeen2009tomography}, thus our MDI QRNG is readily extendable to practical sources, which will be detailed in Section \ref{sec:coh}.

\section{Quantifying randomness} \label{sec:POVM}
After obtaining the two-output POVM set, $\{F_0,F_1\}$ in Eq.~\eqref{Eq:povm}, we need to quantify how much randomness when an input state $\ket{+}$ is fed into the measurement device. Here, we employ the widely-used min-entropy to quantify the randomness.

Given an (even pure) state, the evaluation of the output genuine randomness from a POVM set, $\{F_0,F_1\}$, is not straightforward. A naive approach that the randomness is just the entropy of the outcomes is not working. Consider the case of $F_0=F_1=I/2$, then for any qubit input, both probabilities of outputting 0 and 1 are $1/2$, and hence the outcome entropy is 1. However, Eve could simply output this statistics using a predetermined string (unknown to Alice) without being noticed\footnote{This attack can also be understood as Alice measures one qubit of a maximally entangled pair while Eve measures the other. The ``predetermined" property comes from the fact that Eve can always measure her qubit ahead.}. That is, for this pair of POVMs, no true randomness can be obtained by Alice. Thus, we need to find a way to distinguish classical and quantum randomness. Similar issues are dealt when randomness is used to quantify quantum coherence \cite{Yuan2015Coherence}.

To lower bound the randomness, we should allow Eve to implement the two POVMs in an arbitrary way. Denote Eve's implementation as $\mathcal{D}$ and the randomness corresponding to this implementation as $R(F_0,F_1, \mathcal{D})$. Consider the worst implementation  $\mathcal{D}$ that minimizes $R(F_0,F_1, \mathcal{D})$,
the randomness of the POVM set, $R(F_0,F_1)$, should be
\begin{equation} \label{eq:DecompEntropy}
R(F_0,F_1)= \min\limits_{ \mathcal{D}} R(F_0,F_1, \mathcal{D}).
\end{equation}

As an example of Eve's implementation, Eve can choose a measurement of the following form (the number of terms in the summation below is decided by Eve),
\begin{equation} \eqalign{\label{Eq:generalPOVM}
F_0 &= \sum_i p_i \ket{\psi_i}\bra{\psi_i}, \\
F_1 &= \sum_i p_i \ket{\psi_i^{\bot}}\bra{\psi_i^\bot}+ cI, \\
&\ c+\sum_i p_i =1,
}\end{equation}
which we call \emph{standard decomposition} form. In this decomposition, with a probability of $c$, Eve outputs 1 deterministically, while with probability $1-c$, Eve chooses a set of two-dimensional projection-valued measure (PVM), $\{\psi_i, \psi_i^{\bot}\}$, with a probability distribution $\{p_i\}$, and outputs the measurement outcome 0 or 1. Note that $F_0$ and $F_1$ are fixed due to measurement tomography presented in Sec.~\ref{sec:tomog}.

For a standard decomposition $\mathcal{D}$, we define the randomness when the input is $\ket{+}$ as
\begin{equation} \label{eq:DecompEntropy}
R(F_0,F_1, \mathcal{D})=\sum_i p_i H_{\infty}( |\braket{+}{\psi_i}|^2),
\end{equation}
where $H_{\infty}(p)=-\log_2 \max(p,1-p)$ is the binary min-entropy function. Here is the intuition behind this definition. The total randomness contains two parts: (1) Randomness due to the choice of PVM from the decomposition $\mathcal{D}$. This part contains classical randomness (known to Eve) and thus should be discarded. (2) Randomness associated with each PVM. This part contains real quantum randomness. For a PVM $\{\psi_i, \psi_i^{\bot}\}$, the randomness is quantified by $H_{\infty}( |\braket{+}{\psi_i}|^2)$, as presented in Sec.~\ref{sec:stdrandom}. Note that this definition of randomness also holds for general decompositions.

Although from Alice's point of view, the POVM, $\{F_0,F_1\}$, is two-dimensional, Eve can implement it with arbitrarily large dimension PVMs by adding ancillary systems. Thus, as the first step shown in Sec.~\ref{sec:decomp2dim}, we need to reduce their dimensions down to two. In Sec.~\ref{sec:decompstand}, we reduce a general two-dimensional PVM decomposition to the standard decomposition form in Eq.~\eqref{Eq:generalPOVM}. After that, we evaluate the genuine randomness with the standard decomposition form in Sec.~\ref{sec:stdrandom} and obtain the following theorem. In Sec.~\ref{sec:coll}, we extend this result from individual attacks to collective attacks.

\begin{theorem}
When $\ket{+}$ is fed into the measurement device, described by a POVM set of $\{F_0,F_1\}$ where $F_0=a_1(I+n_x\sigma_x+n_y\sigma_y+n_z\sigma_z)$, the output randomness is given by
\begin{equation}
R(F_0,F_1)=2a_1 H_{\infty}(\frac{1+\sqrt{1-n_y^2-n_z^2}}{2}).
\end{equation}
\end{theorem}

%In this subsection and the next subsection, we will show that actually one of the optimal strategies for Eve is the standard decomposition form presented in Eq.~\eqref{Eq:generalPOVM}, by reducing the most general decomposition strategy of Eve to the standard decomposition form. This is carried out in two steps, first reduce a general measurement decomposition to a decomposition consisting only of two-dimensional projective measurements (Sec. \ref{sec:decomp2dim}) and then reduce a two-dimensional projective measurement decomposition to the standard decomposition form (Sec. \ref{sec:decompstand}).

\subsection{Reduce general measurement to two-dimensional PVM} \label{sec:decomp2dim}
\begin{figure}[hbt]
\centering
\includegraphics[width=10 cm]{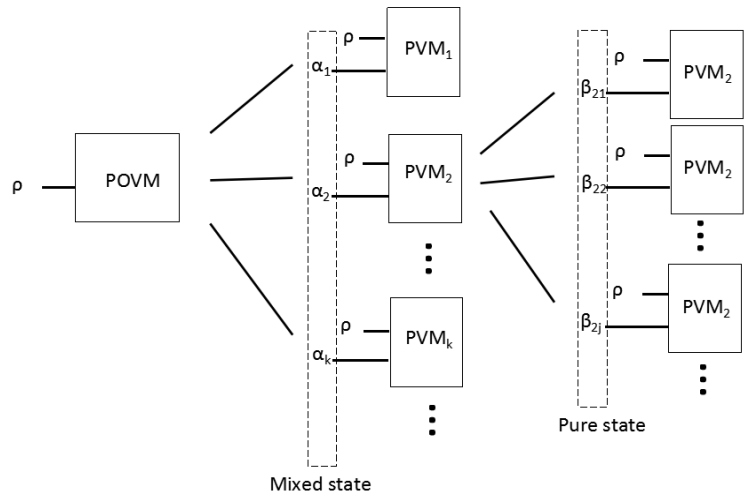}
\caption{POVM decomposition. On the first level of the tree, the POVM on the input $\rho$ is implemented by Eve as an average of projective measurements PVM$_k$ on $\rho$ and a mixed ancilla state $\alpha_k$. On the second level of the tree, each node PVM$_k$ on the first level is further decomposed to PVM$_k$ on $\rho$ and a pure ancilla state $\beta_{kj}$. Note here $\beta_{kj}$ is a decomposition of the mixed state $\alpha_k$. }
\label{Fig:decomp}
\end{figure}

Note that every mixed state is a mixture of pure states. Naturally, we can imagine that every POVM can be decomposed into more basic building blocks, PVMs, as shown in Fig.~\ref{Fig:decomp}. Note that from Alice's view, the measurement is described by a two-dimensional POVM, but she does not know its inner working. While from Eve's view, she is the one who implements POVM with a mixture of different quantum processes, as shown by the branches in Fig.~\ref{Fig:decomp}. Generally, every POVM is a mixture of PVMs on the original state and some ancilla $\alpha_k$  (not necessarily of the same dimension), followed by assigning the outcomes of PVMs to the outcomes of the POVM.

The mixture of PVMs can be implemented by Eve choosing PVM index $k$ according to some random variable. If the random variable is classical, we call it classical adversary. If it is quantum, we call it quantum adversary.

%For the $i$-th round, if Eve reads the $i$-th bit of his advice string to be, say $k$, she then performs PVM$_k$ on the input state with ancilla $a_k$. If Eve is allowed to use a quantum advice $\rho_E$ to determine which device to use, the proof is then generalized from the classical adversary case to the quantum adversary case.

%Suppose the mixture has probability distribution $\{p_k\}$ over the branches, Eve keeps a classical advice string $P_E$ containing the value $k$ with frequency $p_k$ for every $k$. For the $i$-th round, if Eve reads the $i$-th bit of his advice string to be, say $k$, she then performs PVM$_k$ on the input state with ancilla $a_k$. If Eve is allowed to use a quantum advice $\rho_E$ to determine which device to use, the proof is then generalized from the classical adversary case to the quantum adversary case.

In general, each ancilla $\alpha_k$ can be a mixed state, which is decomposed to a spectrum of pure states $\beta_{kj}$. So, a PVM on the input state $\rho$ and the mixed state ancilla $\alpha_k$ can be further decomposed into the PVM on the input state $\rho$ and a statistical mixture of pure state ancillas $\beta_{kj}$, as shown in Fig.~\ref{Fig:decomp}. Thus in the decomposition of a POVM, the ancilla can be assumed to be a pure state $\beta_{kj}$, without loss of generality.
%In all, the randomness of the POVM is a weighted average of the randomness of the composed PVM with pure ancillary states, followed by assignments of outcome values.
Moreover, since a unitary transformation can evolve $\ket{0}$ to any pure ancilla state $\beta_{kj}$, and a unitary transformation can always be encompassed into a PVM, the ancilla can also be viewed to be always in the state of $\ket{0}$. Here, the dimension of $\ket{0}$ can be large.

Now, we can show that decomposing a POVM set into high-dimensional PVMs is equivalent to decomposing into two-dimensional ones. From Eve's point of view, the use of high-dimensional PVMs cannot reduce the output randomness further than using only two-dimensional ones. We first characterize the randomness of a high-dimensional PVM implementation of a POVM set. Then, we decompose the high-dimensional PVM to two-dimensional PVMs, and show that the decomposition cannot increase the output randomness.
%This is proved by the following three steps. The first step is to characterize the randomness of a high-dimensional PVM implementation of a POVM set. The second step is to decompose the high-dimensional PVM to two-dimensional PVMs. The third step is to show the decomposition cannot increase the output randomness.

\begin{figure}[hbt]
\centering
\includegraphics[width=7 cm]{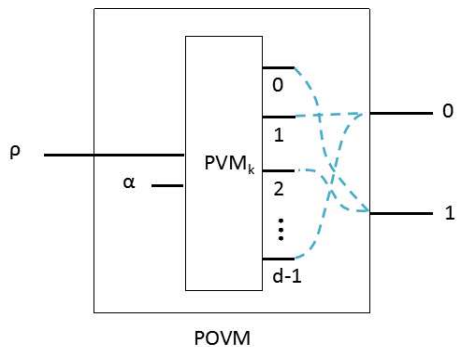}
\caption{An illustration of grouping. For an implementation of a POVM, first a $d$-dimensional PVM$_k$ projects the input state and ancilla to one of its $d$ orthogonal basis and then groups these $d$ outcomes to the two outcomes of the POVM.}
\label{fig:grouping}
\end{figure}

According to Born's rule, the outcomes of PVM is intrinsically random \cite{Yuan2015Coherence}. Now we can quantify the randomness of a high-dimensional PVM. While grouping the output results of PVMs to the ones of the original POVMs, as shown in Fig.~\ref{fig:grouping}, we can view it as a projection onto subspaces, which is still inherently random.

Take the following projection, which is performed as a branch of the decomposition of the original POVM, for an example. It projects 0 to 0, and projects 1 and 2 to 1:
\begin{equation}
 a\ket{0}+ b\ket{1}+ c\ket{2} \rightarrow a\ket{\bar{0}} + \sqrt{b^2 +c^2} \ket{\bar{1}}.
\end{equation}
So according to Born's rule, projecting to the orthogonal subspaces, $\ket{\bar{0}}$ and $\ket{\bar{1}}$, is still random.
In this example,
\begin{equation} \eqalign{
 \mathsf{Prob}(0)&=a^2, \\
 \mathsf{Prob}(1)&=b^2+c^2, \nonumber
}\end{equation}
and so the randomness of this three-dimensional PVM is $H_{\infty}(\mathsf{Prob}(0))$ which is the maximally possible given that the probability of outputting 0 is of value $a^2$. Thus viewing this part as a virtual two-dimensional POVM (note this is different from the original POVM because there are many branches and this is just one of them) and further decompose this POVM to multiple two-dimensional PVMs\footnote{This can always be done by, e.g., the decomposition in Eq.~\eqref{Eq:decomposition} for an arbitrary two-dimensional POVM.} will only decrease the randomness.

More generally, for a general $d$-dimensional PVM, we should also group its outputs to the two outcomes of the original POVM. Suppose the values $v_1,\cdots,v_k$ are projected to 0 ($0\le k\le d$) and $v_{k+1},\cdots, v_d$ are projected 1, then
\begin{equation}
 \sum\limits_{i=0}^{d-1} a_i\ket{i}=\sum\limits_{i=1}^{d} a_{v_i}\ket{v_i} \rightarrow \sqrt{\sum\limits_{i=1}^k a_{v_i}^2}\ket{\bar{0}} + \sqrt{\sum\limits_{i=k+1}^d a_{v_i}^2} \ket{\bar{1}}.
\end{equation}
The randomness is $H_{\infty}(\sum_{i=1}^k a_{v_i}^2)$ and can be similarly reduced through replacing this $d$-dimensional PVM by several branches of two-dimensional PVMs.

%note that a higher-dimensional projective measurement followed by assigning outcome values, can also be expressed as a two dimensional POVM. If we decompose this POVM further into  two dimensional projective measurements, the randomness may be not maximal. This is better than the higher dimensional projective measurement because the higher dimensional one has full randomness. This concludes the proof. In summary, the best strategy of Eve consists of only a mixture of two dimensional projective measurements.

\subsection{Reduce two-dimensional PVM to standard decomposition form} \label{sec:decompstand}
%In the last subsection, we saw that a general measurement decomposition is no better than a two-dimensional projective measurement decomposition. In this subsection, we further reduce a two-dimensional projective measurement decomposition to the standard decomposition form.

The reduction from a two-dimensional PVM decomposition to the standard decomposition form consists of two steps: express the two-dimensional PVM decomposition in a concise form, and then reduce it to the standard decomposition form.

%First note that there will  be two outputs for both a two dimensional POVM $\{F_0,F_1\}$ and a projective measurement PVM. If we construct a mapping between the outputs of the POVM  and PVM, there are in total four cases, as shown in Fig.~\ref{fig:twoPM}.

\begin{figure}[hbt]
\centering
\includegraphics[width=8 cm]{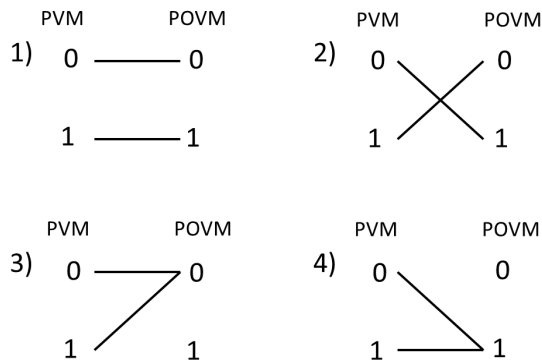}
\caption{Four types of assigning the outcome of PVM (0 or 1) to the outcome of POVM (0 or 1). In the first type, $ \ket{\psi_i}\bra{\psi_i}$ and $ \ket{\psi_i^{\bot}}\bra{\psi_i^\bot}$ contribute to $F_0$ and $F_1$ respectively. In the second type,
  by a change of variable $\ket{\psi_i}=\ket{\phi_i^\bot}$, it is similar to the first case.
  In the third type, $I$ contributes to $F_0$. In the fourth type, $I$ contributes to $F_1$.}
\label{fig:twoPM}
\end{figure}

Recall that in the previous subsection, the outcomes of each $d$-dimensional PVM will be grouped to two values 0 and 1. Take the specific case of $d=2$, there are four types of such grouping, as shown in Fig.~\ref{fig:twoPM}. Denote the two bases of a two-dimensional projective measurement PVM$_i$ as $\ket{\psi_i}$ and $\ket{\psi_i^{\bot}}$, which are orthogonal\footnote{Here the bases of two-dimensional PVM$_i$ are not simply $\ket{0}$ and $\ket{1}$ because different PVM$_i$ have different reference frames. To be consistent, we take the reference frame of the original POVM and PVM$_i$ will accordingly have bases $\ket{\psi_i}$ and $\ket{\psi_i^{\bot}}$.}.
In the first type, $\ket{\psi_i}\bra{\psi_i}$ and $ \ket{\psi_i^{\bot}}\bra{\psi_i^\bot}$ contribute to $F_0$ and $F_1$ respectively. In the second type, $\ket{\psi_i^{\bot}}\bra{\psi_i^\bot}$ and $ \ket{\psi_i}\bra{\psi_i}$  contribute to $F_0$ and $F_1$ respectively. By a change of variable $\ket{\psi_i}=\ket{\phi_i^\bot}$, it is the same as the first case. In the third type, both $ \ket{\psi_i}\bra{\psi_i}$ and $\ket{\psi_i^{\bot}}\bra{\psi_i^\bot}$ contribute to $F_0$. In the fourth type, both $\ket{\psi_i}\bra{\psi_i}$ and $ \ket{\psi_i^{\bot}}\bra{\psi_i^\bot}$ contribute to $F_1$.

By combining all PVMs with assignments of the third type (i.e., $F_0$ will have a term $b_1I$), and combining all PVMs with assignments of the fourth type (i.e., $F_1$ will have a term $b_2I$), a decomposition $\mathcal{D}_1$ has the expression,
\begin{equation} \eqalign{ \label{eq:int}
F_0&=b_1I+ 0 + \sum_{i\ge3} p_i \ket{\psi_i}\bra{\psi_i}, \nonumber\\
F_1&=0  + b_2 I  +\sum_{i\ge3} p_i \ket{\psi_i^{\bot}}\bra{\psi_i^\bot},\nonumber\\
&b_1+b_2+\sum_{i\ge3} p_i=1,
}\end{equation}
where the summation comes from PVMs with assignments of the first type and the second type.

Next, we prove it can be reduced to the standard decomposition form in the sense that the value of $R(F_0,F_1)$ will not change when restricting the minimization over the standard decomposition form.
Take $c=b_2-b_1$, we obtain a decomposition $\mathcal{D}_2$, which is equivalent to $\mathcal{D}_1$.
\begin{equation} \eqalign{ \label{eq:form1}
F_0&=b_1\ket{+}\bra{+}+b_1\ket{-}\bra{-}+ \sum_{i\ge3} p_i \ket{\psi_i}\bra{\psi_i},\nonumber\\
F_1&=b_1\ket{-}\bra{-}+b_1\ket{+}\bra{+}  +\sum_{i\ge3} p_i \ket{\psi_i^{\bot}}\bra{\psi_i^\bot}+ cI,\nonumber\\
&2b_1+c+\sum_{i\ge3} p_i=1,
}\end{equation}
Let $\ket{\psi_1}=\ket{+}$ (thus $\ket{\psi_1^{\bot}}=\ket{-}$), $\ket{\psi_2}=\ket{-}$ (thus $\ket{\psi_2^{\bot}}=\ket{+}$) and $p_1=p_2=b_1$, then the decomposition $\mathcal{D}_2$ is in the standard decomposition form Eq.~\eqref{Eq:generalPOVM}.

Finally, we just need to prove that
\begin{equation}
R(F_0,F_1,\mathcal{D}_1)=R(F_0,F_1,\mathcal{D}_2).
\end{equation}
On one hand, $F_0=I$ and $F_1=0$ means that the output is always 0 and there is no randomness. On the other hand,
$H_{\infty}(\braket{+}{+})=H_{\infty}(\braket{+}{-})=0$ also gives no randomness. Thus the difference between the two decompositions gives no randomness and thus they are equal in all cases.

\subsection{Minimization of standard decomposition form}  \label{sec:stdrandom}
From the previous two subsections, we conclude that without loss of generality, the strategy of Eve can be restricted to the standard decomposition form. In this subsection, we allow Eve to choose the best strategy within the standard decomposition form. Recall that in this case, the randomness measure for the POVM can be expressed as
\begin{equation}
R(F_0,F_1)=\min_{p_i,\ket{\psi_i}}\sum_i p_i H_{\infty}( |\braket{+}{\psi_i}|^2),
\end{equation}

According to Eq.~\eqref{Eq:generalPOVM}, a simple example of decomposition of the POVM can be given by,
\begin{equation} \eqalign{\label{Eq:decomposition}
F_0&=a_1(1-|\vec{n}_1|)I+a_1(|\vec{n}_1|I+\vec{n}_1\cdot \sigma), \nonumber \\
F_1&=a_2(1-|\vec{n}_2|)I+a_2(|\vec{n}_2|I+\vec{n}_2\cdot \sigma).
}\end{equation}
whose randomness property and relation to the standard decomposition form is proven in \ref{app:exam}. In particular, note that $a_1(|\vec{n}_1|I+\vec{n}_1\cdot \sigma)$ and $a_2(|\vec{n}_2|I+\vec{n}_2\cdot \sigma)$ are a set of PVMs because $a_1\vec{n}_1+a_2\vec{n}_2=0$. Thus one can obtain a random measurement outcome for this decomposition. However, this may not be the optimized decomposition for Eve, because the output randomness for this decomposition will be larger than $R(F_0,F_1)$. Then following some previous work, which utilizes a general decomposition to quantify randomness \cite{decomp}, we try to obtain an accurate expression of the minimum randomness $R(F_0,F_1)$ corresponding to an optimized decomposition of the POVM.

A general expression of a mixed state can be written as:
\begin{equation} \eqalign{\label{Eq:generalmixedstate}
\rho &=\sum_i q_i \ket{\varphi_i}\bra{\varphi_i},\nonumber\\
&=\frac{(I+n_x\sigma_x+n_y\sigma_y +n_z\sigma_z)}{2}.
}\end{equation}
When performing a measurement on the bases $\{ \ket{+}, \ket{-}\}$, the outcome randomness can be expressed as
\begin{equation}\label{Eq:randomnessrho}
R(\rho)=\min_{q_i,\ket{\varphi_i}}[H_{\infty}(q_i)+\sum\limits_iq_i H_{\infty}(|\braket{\varphi_i}{+}|^2)],
\end{equation}
where the first term $H_{\infty}(q_i)$ represents the classical randomness originating from the probability distribution of $q_i$, and it should be discarded in the following analysis. Thus, the net quantum randomness output is given by
\begin{equation}\label{Eq:Qrandomnessrho}
R(\rho)=\min_{q_i,\ket{\varphi_i}}\sum\limits_iq_i H_{\infty}(|\braket{\varphi_i}{+}|^2).
\end{equation}

When performing the POVM given in Eq.~\eqref{Eq:generalPOVM} on an input state $\ket{+}$, since the term $cI$ generates no randomness, the output randomness has a similar form
\begin{equation}\label{Eq:randomnessPOVM}
R(F_0,F_1)=\min_{p_i,\ket{\psi_i}} \sum\limits_i p_i H_{\infty}(|\braket{+}{\psi}|^2).
\end{equation}
%where the classical randomness $H_{\infty}(p_i)$ should also be discarded.

The bases in the PVM $\{\ket{\psi}, \ket{\psi^\bot}\}$ and an arbitrary pure state $\ket{\phi}$ have a natural duality. That is, the probability of projecting $\ket{\phi}$ on $\ket{\psi}$ is equal to that of projecting $\ket{\psi}$ on $\ket{\phi}$:
\begin{equation}
|\braket{\psi}{\phi}|^2=|\braket{\phi}{\psi}|^2.
\end{equation}
Then we can easily find the quantum randomness in Eq.~\eqref{Eq:Qrandomnessrho} and Eq.~\eqref{Eq:randomnessPOVM} are the same.

In addition, the measurement basis $\ket{\psi_i}\bra{\psi_i}$ has a pure state form
\begin{equation}\label{Eq:purestateformofpsii}
\ket{\psi_i}\bra{\psi_i}=\frac{I+n_i\cdot \sigma}{2}.
\end{equation}
Combining Eq.~\eqref{Eq:povm} and Eq.~\eqref{Eq:generalPOVM} we can get
\begin{equation}
\sum_i p_i=2a_1.
\end{equation}
Then if we let
$p'_i=p_i/2a_1$,
the quantum randomness $R(F_0,F_1)$ can be rewritten as
\begin{equation} \label{eq:randrhosim}
R(F_0,F_1)=2a_1\min_{p'_i,\ket{\psi_i}} \sum\limits_i p'_i H_{\infty}(|\braket{\psi_i}{+}|^2).
\end{equation}

According to related study to quantify randomness for a mixed state and PVM \cite{PhysRevA.75.032334, Yuan2015Coherence}, the mixed state randomness in Eq.~\eqref{Eq:Qrandomnessrho} can be expressed as
\begin{equation}
R(\rho)= H_{\infty}(\frac{1+\sqrt{1-n_y^2-n_z^2}}{2}).
\end{equation}
Thus Eq.~\eqref{eq:randrhosim}, as well as Eq.~\eqref{Eq:randomnessPOVM}, can be simplified to
\begin{equation}\label{eq:finalrandom}
R(F_0,F_1)=2a_1 H_{\infty}(\frac{1+\sqrt{1-n_y^2-n_z^2}}{2}).
\end{equation}
One can see that, as long as $n_y$ or $n_z$ is nonzero, $R(F_0,F_1)$ is always positive. Note that the choice of $\ket{+}$ is not compulsory. Other input states can be used as randomness generation by a simple rotation of the reference frame. Take $\ket{0}$ for example, the randomness of the outcome corresponding to this new input state is $R(F_0,F_1)=2a_1 H_{\infty}((1+\sqrt{1-n_x^2-n_y^2})/2)$.

\subsection{From individual attack to collective attack} \label{sec:coll}
Now, we have showed the quantification of output randomness under individual attacks. For a collective attack, since Eve can perform independent different attacks to each run. That is, for the $i$th round ($1\le i\le n$), she performs POVM$_i$, thus the total output randomness is
\begin{equation}
\sum\limits_{i=1}^n R(\textrm{POVM}_i).
\end{equation}

If the function $R$ is convex, we have
\begin{equation}
\sum\limits_{i=1}^n R(\textrm{POVM}_i) \ge n R(\frac{\sum_{i=1}^n\textrm{POVM}_i}{n}),
\end{equation}
where the expression in the bracket on the right hand side is exactly the tomography result. So, in order to generalize individual attacks to collective attacks, it suffices to examine the convexity property of the randomness quantification Eq.~\eqref{eq:finalrandom}, as shown in \ref{app:convex}. Thus, our randomness quantification Eq.~\eqref{eq:finalrandom} holds against collective attacks.

%Finally, the discussion on extending collective attack to coherent attack will be deferred to the last section.

\section{Statistical fluctuation}
\label{sec:stat}
The above analysis assumes that the protocol has an infinite number of runs, such that the parameters can be accurately estimated. However in practice, protocols are only allowed to run for a finite amount of time, which results in imperfect tomography due to statistical fluctuations. Thus in this section, we take account of the finite-size effect by bounding the key parameters $a_1$, $a_1n_x$, $a_1n_y$, $a_1n_z$ in Eq.~\eqref{eq:tom}, using the techniques in QKD \cite{Fung:Finite:2010}. Whether to use the upper bound or the lower bound of the parameters, depends on which gives the minimum randomness output according to our previous analysis. This will give the most conservative estimate on the output randomness.

In a test run, Alice sends one of the four states $\rho_1=I-\sigma_z$, $\rho_2=I+\sigma_x$, $\rho_3=I+\sigma_y$, $\rho_4=I+\sigma_z$ and obtains the probabilities of outputting $0$, denoted by $e_{x1}, e_{x2},e_{x3}, e_{x4}$ that correspond to their asymptotic values $a_1-a_1n_z$, $a_1+a_1n_x$,  $a_1+a_1n_y$, $a_1+a_1n_z$, respectively.
%From now on, we denote $N_{xi},\ i=1,2,3,4$ for the number of runs for each of the four states and $e_{xi}$ for the corresponding four values.
After the protocol finishes, the number of test runs with input $\rho_i$ is denoted as $N_{i}$, $i=1,2,3,4$.

Let $N_{0}$ denote the number of non-test runs. Recall that in each non-test run, Alice sends $\rho_2=\ket{+}\bra{+}$. Let  $e_{zi}$ be the probability of outputting $0$ if the input of the non-test runs were $\rho_i$ instead.
 Define the bound,
\begin{equation} \label{eq:eptheta}
e_{zi}\le e_{xi}+\theta, \quad i=1,2,3,4
\end{equation}
where $\theta$ is the deviation due to statistical fluctuations.

Following the random sampling results of Fung et al.~\cite{Fung:Finite:2010}, we can bound the quantity $e_{z1}$ when Eq.~\eqref{eq:eptheta} fails,
\begin{equation} \eqalign{
\varepsilon_\theta &= \textsf{Prob}(e_{z1}>e_{x1}+\theta)  \\
&\le \frac{\sqrt{N_1+N_0}}{\sqrt{N_1 N_0 e_{x1}(1-e_{x1})}}2^{-(N_1+N_0)\xi_1(\theta)}, \nonumber
}\end{equation}
where $\xi_1(\theta)= H(e_{x1}+N_0\theta/(N_0+N_1))-[N_1 H(e_{x1})+N_0H(e_{x1}+\theta)]/(N_0+N_1)$. Here $H(p)= -p\log p-(1-p)\log (1-p)$ is the binary Shannon entropy function. Note that in an unlikely event when $e_{x1}=0$, one should re-derive the failure probability or simply replace $e_{x1}$ with a small value, say, $1/N_1$.

Note that the original random sampling trick is applied on variables between $[0,1]$. However, the range of $e_{zi}$ is $[-1,1]$ for $i=2,3,4$.  This requires a normalization which scales from $[-1,1]$ to $[0,1]$. This normalization transforms $y$ to $y'=(1+y)/2$ which yields
\begin{equation} \eqalign{
\varepsilon_\theta &= \textsf{Prob}(e_{zi}>e_{xi}+\theta) \quad \forall i=2,3,4   \\
&\le \frac{4\sqrt{N_i+N_0}}{\sqrt{N_i N_0 (1+e_{xi})(1-e_{xi})} }2^{-(N_i+N_0)\xi_i(\theta)},\nonumber
}\end{equation}
where $\xi_i(\theta)=H((1+e_{xi})/2+N_0\theta/(N_0+N_i))-[N_i H((1+e_{xi})/2)+N_0H((1+e_{xi})/2+\theta)]/(N_0+N_i)$.

Practically, we can let the failure probability $\varepsilon_\theta$ to be a small number for certain applications, say $2^{-100}$. Once the upper bound of $\varepsilon_\theta$ is fixed, there is a trade-off between $N_i/(N_0+N_i)$, $\theta$ and the ratio of the final random bit length over the raw data size $R(F_0,F_1)$. In addition, a suitable $N_i$ can be chosen to optimize $R(F_0,F_1)$.

Also note that the input randomness is on the order of $\log N_0$ to achieve a desired small failure probability, while the output randomness is on the order of $N_0$, thus an exponential expansion of randomness is achieved.

\section{From single photon source to coherent source} \label{sec:coh}
In practice, a weak coherent state photon source (highly attenuated laser) is widely used as an imperfect single photon source. To make our MDI QRNG scheme practical, we need to use a coherent light as the trusted source. This change introduces two obstacles in analysis. One is that the input states are changed in tomography.
% Though coherent state can be used to tomography the photon number states, here our goal is different in that we need to tomography a two dimensional POVM instead of an infinite dimensional one.
The other is the final output randomness is different. Since the intensity of the source can be used to estimate the single photon component emitted from the source, we can bound the output randomness with an ``imperfect" tomography.

%Fortunately, when the source is a coherent state, we could apply the technique by Lloyd et al \cite{PhysRevX.4.011016} and the result is similar to QKD without decoy state.
For a coherent state with a mean photon number $\mu$, a phase randomization procedure transforms a superposition of Fork state into a mixture. In other words, the final state can be divided into three components, vacuum, single photon, and multi-photon. Since these three parts are orthogonal, they can be treated as different channels separately. By controlling the intensity $\mu$ low enough, the multi-photon component can be suppressed. We prove a lower bound on the randomness of our MDI QRNG with a coherent state source, using a series of relaxations.

\begin{figure}[hbt]
\centering
\includegraphics[width=10 cm]{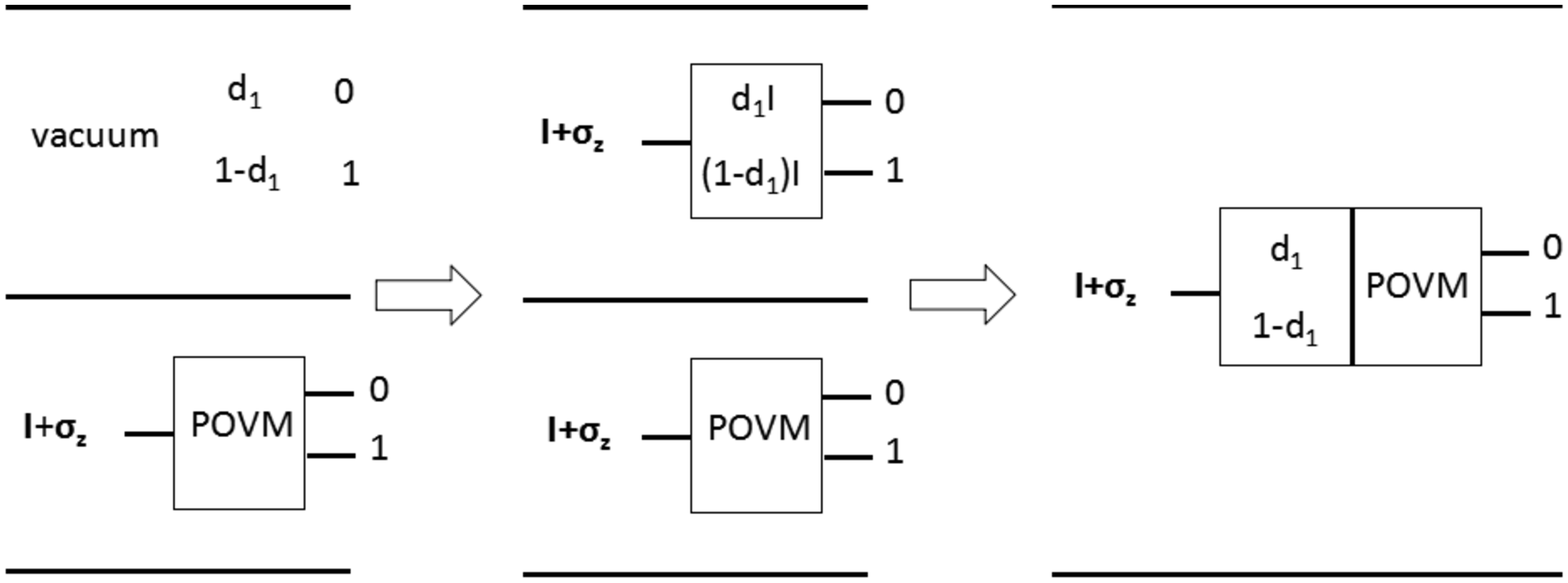}
\caption{Combining the channel of the vacuum component and the single photon component. Note that after combination, the randomness can only decrease, giving a lower bound on the original channels.}
\label{fig:vac}
\end{figure}

As for the vacuum component, in the worst case scenario, we assume the adversary Eve is able to determine the outcomes ahead, and hence no true randomness can be generated. As shown in Fig.~\ref{fig:vac}, the measurement is equivalent to a virtual qubit measurement with $F_0=d_1 I$ and $F_1= (1-d_1)I$ on any qubit state input.

With these preparations at hand, we now can perform tomography on the qubit-POVM with a coherent state. Denote the POVM of the single photon component to be $F_0'=  d_1' I + d_2'\sigma$, $F_1'=(1-d_1')I - d_2' \sigma$. Since the proportion of the vacuum and the single photon component are $e^{-\mu}$ and $\mu e^{-\mu}$, we can combine the POVM of the single photon with the virtual POVM on the vacuum,
\begin{equation} \eqalign{
F_0''&=d_1 I  e^{-\mu} + (d_1' I + d_2'\sigma)\mu e^{-\mu} \\
F_1''&=(1-d_1) I e^{-\mu} + ((1-d_1' )I - d_2'\sigma)\mu e^{-\mu}, \nonumber
}\end{equation}
 as shown in Fig.~\ref{fig:vac}. Here the combined channel will have a proportion that is the sum of the proportion of single photon and vacuum in the original channels, which is $(1+\mu)e^{-\mu}$.

We now verify such a combination will not be an overestimate on the output randomness. Originally the actual randomness comes from each separate component, which corresponds to $F_0,F_1$ for the vacuum and $F_0',F_1'$ for single photon.
Since the output randomness of $F_0, F_1$ is independent of its qubit input, without loss of generality, the input of $F_0,F_1$ can be set to the single photon component input. For example, as illustrated in the middle part of Fig.~\ref{fig:vac}, since the qubit input to the single photon component is $I+\sigma_z$, the input of the virtual measurement $F_0,F_1$ is also set to $I+\sigma_z$.
Recall that the randomness measure is the minimum over all decompositions. Since the decomposition $F_0'=F_0e^{-\mu}+F_0'\mu e^{-\mu}, F_0'=F_0e^{-\mu}+F_0'\mu e^{-\mu}$ is also a decomposition of a combined POVM and this decomposition corresponds to exactly the sum of the original randomness of vacuum and single photon channels,  the randomness measure of the combined POVM can serve as a lower bound on the original randomness. Hence, using this combined POVM will not overestimate the output randomness.

In summary, vacuum component and single photon component can be combined as one source to generate randomness and previous analysis in Sec.~\ref{sec:POVM} still applies. That is, for randomness generation purpose, both vacuum state and single-photon state can be regarded as an ideal qubit state. This is similar to QKD, where vacuum state can also be used to generate secure keys \cite {Lo:Vacuum:2005}.

Now we need to take multi-photons components into account. We consider the worst case scenario \cite{GLLP:2004} where multi-photon components do not contribute to randomness generation.

In addition, multi-photon states have the effect of making the tomography imperfect. We conservatively assume multi-photon states will always lead to a tomography outcome which minimizes the output randomness.
%Next, we extract the information of $I$, $I+\sigma_x$, $I+\sigma_y$, $I+\sigma_z$ from the above statistics. If there are no multi-photon components, this can be done perfectly.
%There are $e^{\mu}-1-\mu$ multi-photons,
%which we take the worst guarantee that part of the 1 are actually 0 by faking those parts.
In order to make the randomness smaller, according to Eq.~\eqref{eq:finalrandom}, Eve should make $a_1$, $n_x$ and $n_y$ smaller. Considering the multi-photons components, after POVM on the new input state $\tau_i,(i=1,2,3,4)$,
 the constrains on the probabilities of the output 0 for $\tau_i$ are respectively
\begin{equation} \eqalign{
0&\le \textsf{Prob}(0|\tau_1)-a_1(1+\mu) e^{-\mu}\le 1-e^{-\mu}-\mu e^{-\mu} \nonumber\\
&(a_1+a_1n_x)(1+\mu) e^{-\mu}\le  \textsf{Prob}(0|\tau_2)  \label{eq:constraint}\\
&(a_1+a_1n_y)(1+\mu) e^{-\mu} \le \textsf{Prob}(0|\tau_3)\nonumber\\
&(a_1+a_1n_z)(1+\mu) e^{-\mu}\le \textsf{Prob}(0|\tau_4)\nonumber,
}\end{equation}
%Recall $a_1$ is quite small when $\mu$ is small, with a typical value of $O(\mu\eta)$.
where equalities hold when the multi-photon component does not yield the result of 0 for the last three inequalities. So the bounds of the parameters can be estimated through experimentally obtaining $\textsf{Prob}(0|\tau_i)$, $(1\le i\le 4)$.

Then we estimate the randomness from the vacuum and single photon component, which are combined as shown in Fig.~\ref{fig:vac}. Thus after calculating randomness of the tomographies POVM with input state $(I+\sigma_x)/2$, we multiply by a factor of $(1+\mu) e^{-\mu}$, which is the proportion of the single photon and the vacuum components,
\begin{equation}\label{eq:finalrandomwithmu}
 R(F_0,F_1)\ge\max\limits_\mu  \frac{2a_1(1+\mu)}{e^{\mu}} H(\frac{1+\sqrt{1-n_z^2-n_y^2}}{2}),
\end{equation}
where the parameters are constrained by Eq.~\eqref{eq:constraint}.

\begin{figure}[hbt]
\centering
\includegraphics[width=10 cm]{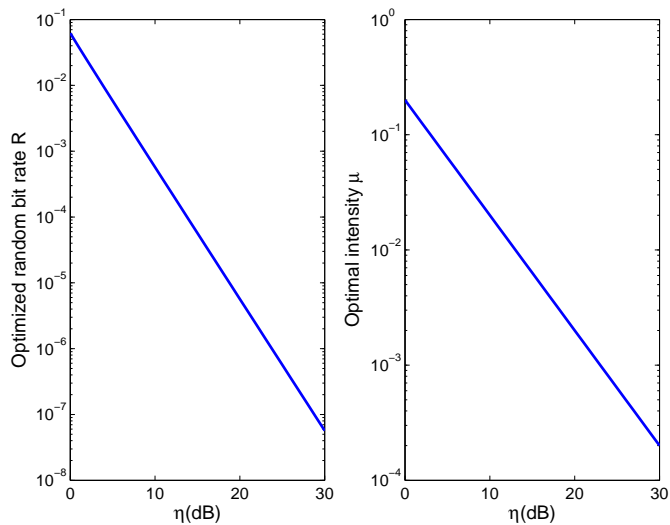}
\caption{Random bit rate $R$ with a coherent state source of an average photon number $\mu$. The left figure shows the dependency of the optimized bit rate on the transmission loss. The right figure shows the average photon number $\mu$ corresponding to the optimal bit rate.}
\label{fig:coh}
\end{figure}

We simulate a typical experiment setup to examine the dependency of random bit rate $R$ on the total transmittance $\eta$. In this setup, a coherent laser with intensity $\mu$ and polarization $\ket{+}$ sends pulses to a measurement apparatus that performs $Z$ basis measurement with low efficiency detectors. The results are shown in Fig.~\ref{fig:coh}, with the simulation details in \ref{app:sim}.

In practice, the laser intensity can be adjusted to optimize the performance. Thus in the simulation, we numerically optimize the laser intensity $\mu$ to maximize the random bit rate $R$. By the simulation, the optimal intensity of the coherent state $\mu$ is approximately proportional to $\eta$ ($\mu \approx 0.2\eta$), which can be seen from the right panel of Fig.~\ref{fig:coh}.

The logarithm of the optimal random bit rate is approximately proportional to the logarithm of $\eta$, as can be seen from the left panel of Fig.~\ref{fig:coh}. Moreover, by examining the figure more carefully, the random bit rate decreases by $10^6$ when the transmittance $\eta$ decreases from $0$ db to $30$ db. Thus the optimal random bit rate $R$ scales quadratically with $\eta$.

These scalings are similar to the early analysis of QKD \cite{GLLP:2004}, where the optimal intensity is also linear with the transmittance and the key rate is quadratic with the transmittance. In the development of QKD, the decoy state technique has increased the key rate to be linear with the transmittance \cite{MXFPhD}. It would be interesting to explore whether similar ideas can be used to improve the random bit rate in our protocol.

With a typical 100 MHz repetition rate laser and a typical total transmittance value $\eta=10\%$, the simulation shows that the random bit rate is over $5\times 10^4$ bit/sec, which is five magnitudes higher than the current record of DIQRNG, 0.4 bit/sec \cite{PhysRevLett.111.130406}.

% Do a simulation which makes the result more convincing. X axis mu, Y axis min-entropy, other parameters perfect

\section{Conclusion}
\label{sec:concl}
In summary, we have proposed a measurement-device-independent QRNG. Our QRNG works when the detectors have low efficiency and have arbitrary imperfections. In contrast to MDI-QKD and MDI-EW, our protocol does not need space-like separation, which can be intuitively explained by the fact that one should perform error correction and privacy amplification in QKD, while one only needs to perform privacy amplification in QRNG.
There are two possible implementations of our scheme, either by using a single photon source or by using a coherent state. The former has higher random bit rate while the latter is more practical.

For future work, it would be interesting to extend the analysis to coherent attack. Intuitively, the best coherent attack is usually just the collective attack. Since our protocol is permutation invariant, that is, the order of different runs can be arbitrarily changed, we can extend the analysis from collective attack to coherent attack possibly by applying the Post-Selection principle \cite{christandl2009postselection}, which may give a moderate increase on the security parameter. Or we can possibly use the work of Miller and Shi \cite{Miller15} to extend from a classical adversary to a quantum adversary, which is essentially the difference between collective attack and coherent attack.

\section*{Acknowledgements}
The authors acknowledge insightful discussions with X.~Yuan and Z.~Zhang. This work was supported by the 1000 Youth Fellowship program in China.

\appendix

\section{Proof that Eq.~\eqref{Eq:decomposition} is of standard decomposition form}
\label{app:exam}
In this section we show that Eq.~\eqref{Eq:decomposition} is a meaningful special case of Eq.~\eqref{Eq:generalPOVM} and analyse the randomness generation of each term of the POVM in Eq.~\eqref{Eq:decomposition}.

In Eq.~\eqref{Eq:decomposition}, if we let $c_1=a_1(1-|\vec{n}_1|),c_2=a_2(1-|\vec{n}_2|),c=c_2-c_1$, then Eq.~\eqref{Eq:decomposition} can be rewritten as
\begin{equation} \eqalign{
F_0&=c_1(\ket{0}\bra{0}+\ket{1}\bra{1})+a_1(|\vec{n}_1|I+\vec{n}_1\cdot \sigma),\\
F_1&=c_1(\ket{1}\bra{1}+\ket{0}\bra{0})+cI+a_2(|\vec{n}_2|I+\vec{n}_2\cdot \sigma) \nonumber
}\end{equation}
Comparing with Eq.~\eqref{Eq:generalPOVM}, we can see that $c_1(\ket{0}\bra{0}+\ket{1}\bra{1})$ and $c_1(\ket{1}\bra{1}+\ket{0}\bra{0})$ are two terms of
$\sum p_i \ket{\psi_i}\bra{\psi_i}$ and $\sum p_i \ket{\psi_i^{\bot}}\bra{\psi_i^\bot}$, respectively. According to Eq.~\eqref{eq:coefficientsrelation}, we have
\begin{equation}
a_1|\vec{n}_1| = a_2|\vec{n}_2|
\end{equation}
\begin{equation}
a_1 \vec{n}_1 \cdot \sigma = -a_2 \vec{n}_2 \cdot \sigma
\end{equation}
  Therefore the rest part $a_1(|\vec{n}_1|I+\vec{n}_1\cdot \sigma)$ and $a_2(|\vec{n}_2|I+\vec{n}_2\cdot \sigma)$ have the same coefficients, which compose the other terms of $\sum p_i \ket{\psi_i}\bra{\psi_i}$ and $\sum p_i \ket{\psi_i^{\bot}}\bra{\psi_i^\bot}$.
\begin{figure}[hbt]
\centering
\includegraphics[width=10 cm]{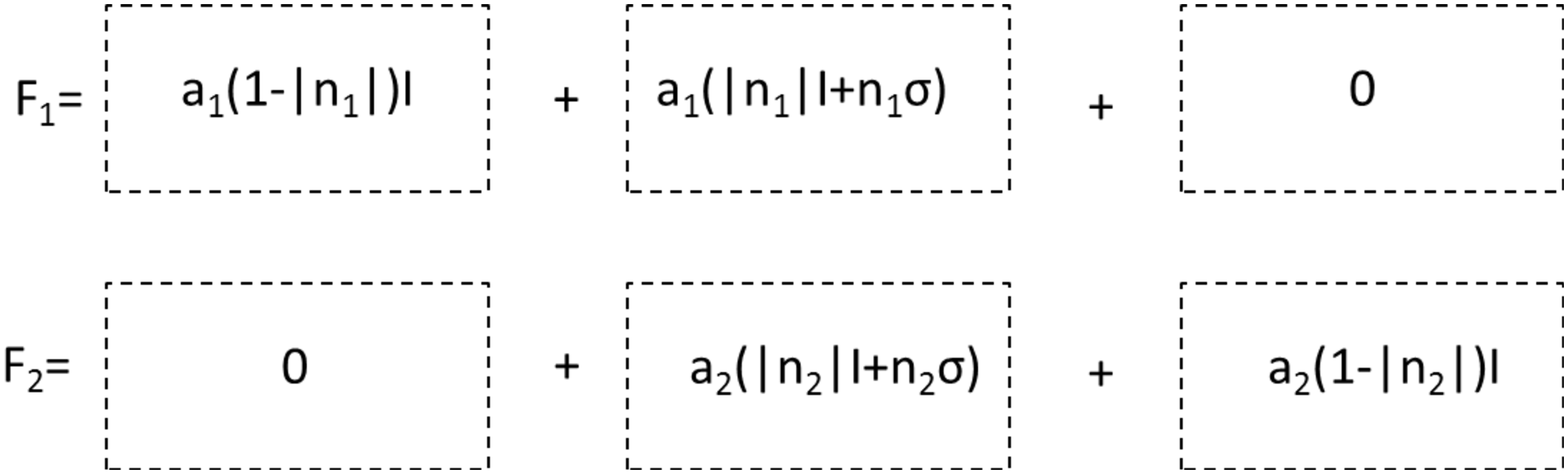}
\caption{The first and third dashed boxes have no contribution to the randomness, while the second one does.}
\label{Fig:formula}
\end{figure}

For such a decomposition, considering an arbitrary input state in $\ket{0},\ket{1}, \ket{+}, \ket{+i}$, we can easily check that the output randomness only originate from the term $a_1(|\vec{n}_1|I+\vec{n}_1\cdot \sigma)$ and $a_2(|\vec{n}_2|I+\vec{n}_2\cdot \sigma)$, as shown in Fig.~\ref{Fig:formula}, which is consistent with our previous results.

\section{Convexity of Eq.~\eqref{eq:finalrandom}} \label{app:convex}
We notice that $R$ in Eq.~\eqref{eq:finalrandom} is a linear function of $a_1$, thus it is convex with respect to $a_1$. For $a_1n_x$, since it does not appear in Eq.~\eqref{eq:finalrandom}, the convexity with respect to $a_1n_x$ also holds. For $a_1n_z$ and $a_1n_y$, due to the symmetry, we just need to check for one of them, and denote $z=a_1n_y$. A direct calculation of the second order derivatives of $z$ on $R$ gives
\begin{equation} \eqalign{
&\quad \frac{\partial^2 R(F_0,F_1)}{\partial z^2}\nonumber \\
&=\frac{\partial^2 2a_1 H_{\infty}((1+\sqrt{1-(z/a_1)^2-n_z^2})/2)}{\partial z^2}\nonumber\\
&=\frac{Cz^2/a_1^3}{(1-(z/a_1)^2-n_z^2)(1+\sqrt{1-(z/a_1)^2-n_z^2})^2}\nonumber \\
& \quad +\frac{Cz^2/a_1^3}{(1-(z/a_1)^2-n_z^2)^{2/3}(1+\sqrt{1-(z/a_1)^2-n_z^2})}\nonumber \\
& \quad +\frac{C/a_1}{\sqrt{1-(z/a_1)^2-n_z^2}(1+\sqrt{1-(z/a_1)^2-n_z^2})} \nonumber\\
&\ge 0,
}\end{equation}
where $C=1/\ln 2$.
Since the second order derivative is positive, the convexity holds for $z=a_1n_y$.

There is another way to prove the convex property of $R$.
%% expand
Recall that the randomness measure $R$ is obtained by a minimization over all possible decomposition of a POVM. For such a convex roof measure,  since the best decomposition of POVM$_i$ ($1\le i \le n$),
$\{p_{ij}, \ket{ \psi_{ij}} \}_{j=1,\cdots, m_i}$
also constitutes a decomposition of $\sum \textrm{POVM}_i/n$,
$\{p_{ij}/n, \ket{ \psi_{ij}} \}_{i=1,\cdots, n, j=1,\cdots, m_i}$,
we have
\begin{equation} \eqalign{
\frac{1}{n}\sum\limits_{i=1}^n R(\textrm{POVM}_i) & = \frac{1}{n}\sum\limits_{i=1}^n  ( \sum\limits_j p_{ij} H_{\infty}(|\braket{0}{\psi_{ij}}|^2)) \nonumber \\
& = \sum\limits_{ij}  \frac{p_{ij}}{n} H_{\infty}(|\braket{0}{\psi_{ij}}|^2) \\
& \ge R(\frac{\sum_{i=1}^n\textrm{POVM}_i}{n}), \nonumber
}\end{equation}
thus the convexity holds.

\section{Simulation} \label{app:sim}
Here are details for the simulation model. A phase randomization procedure transforms a coherent state to a mixture of Fock states. With a mean photon number $\mu$, the probabilities of vacuum component, single photon component and multi-photon component are respectively $e^{{-\mu}}$, $\mu e^{{-\mu}}$, $1-e^{{-\mu}}-\mu e^{{-\mu}}$. Considering the $Z$ basis measurement on such a input mixed state, assuming a no-detection event to be mapped into output 1, the probability of output 0 is given by
\begin{equation} \eqalign{
\textsf{Prob}(0) &=\textsf{Prob}(0|vacuum)e^{{-\mu}} \\
&+\textsf{Prob}(0|single photon)\mu e^{{-\mu}} \nonumber\\
&+\textsf{Prob}(0|multiphoton)(1-e^{{-\mu}}-\mu e^{{-\mu}})\nonumber
}\end{equation}
In experiments, the polarization of the single photon component can be adjusted into the following four states $=I$, $I+\sigma_x$, $I+\sigma_y$, $I+\sigma_z$. Setting $\textsf{Prob}(0|multiphoton)$ to be 0 and 1, the bound of $\textsf{Prob}(0)$ of the corresponding four input coherent state $\tau'_i$ $(i=1,2,3,4)$ are
  %%  Perhaps explain in more detail here
\begin{equation} \eqalign{
\eta \mu  e^{-\mu}/2 \le \textsf{Prob}(0|\tau'_1)\le & \eta \mu e^{-\mu}/2 +(e^{\mu}-1-\mu) e^{-\mu}  \nonumber\\
\eta \mu  e^{-\mu}/2 \le \textsf{Prob}(0|\tau'_2) \le & \eta \mu e^{-\mu}/2 +(e^{\mu}-1-\mu) e^{-\mu}   \nonumber \\
\eta \mu  e^{-\mu}/2  \le \textsf{Prob}(0|\tau'_3) \le &  \eta \mu  e^{-\mu}/2+(e^{\mu}-1-\mu) e^{-\mu}   \\
\eta \mu  e^{-\mu}  \le \textsf{Prob}(0|\tau'_4)\le &  \eta \mu  e^{-\mu}+(e^{\mu}-1-\mu) e^{-\mu} \nonumber
}\end{equation}

Comparing with Eq.~\eqref{eq:constraint}, we can easily obtain the constrains on parameters $a_1$, $n_y$, and $n_z$. According to Eq.~\eqref{eq:finalrandomwithmu}, for an arbitrary set of $a_1$, $n_y$, and $n_z$, we can find an optimal $\mu$ to maximize the final randomness $R(F_0,F_1)$.  Then $R(F_0,F_1)$ can be calculated based on its monotonicity and an optimal $\mu$.

The result of our simulation model is shown in Fig.~\ref{fig:coh}. We can easily check that the final output randomness will be positive.
%When the input state is a coherent state with a mean photon number $\mu$, the lower bound of $R(F_0,F_1)$ decreases with an increasing $\eta$, but will never decrease to $0$, that is, we can always obtain a positive randomness after the measurement of Eve. The optimal $\mu$ has a similar relation with $\eta$. %%%% Then cite the randomness quantification in the main text.  Numerical  optimization over mu

\section*{References}

%%%%%%%%%%%%%%%%%%%%%%%%%%%%%%%%%%%%%%%%
% choose a style
%\bibliographystyle{ieeetr}
%\bibliographystyle{unsrt}
%\bibliographystyle{apsrev4-1}
%\bibliographystyle{discrete}
 \bibliographystyle{iopart-num}
%%%%%%%%%%%%%%%%%%%%%%%%%%%%%%%%%%%%%%%%

\bibliography{Biblimdirng}

\end{document}